\documentclass[aps,prl,twocolumn,amsmath,amssymb,showpacs, superscriptaddress,notitlepage,longbibliography,10pt]{revtex4-1}

\usepackage[colorlinks=true,linkcolor=blue,anchorcolor=red,citecolor=blue, urlcolor=blue]{hyperref}
\usepackage{bm}
\usepackage{graphicx}
\usepackage{physics}
\usepackage{color}
\usepackage{multirow}
\usepackage{extarrows}
\usepackage{array}
\usepackage{appendix}

\usepackage{cases}
\usepackage[normalem]{ulem}

\begin{document}

\title{Coulomb Drag in Altermagnets}

\author{Hao-Jie Lin$^\dag$}
\affiliation{Department of Physics, State Key Laboratory of Quantum Functional Materials, and Guangdong Basic Research Center of Excellence for Quantum Science, Southern University of Science and Technology (SUSTech), Shenzhen 518055, China}
\affiliation{Quantum Science Center of Guangdong-Hong Kong-Macao Greater Bay Area (Guangdong), Shenzhen 518045, China}

\author{Song-Bo Zhang$^\dag$}
\affiliation{Hefei National Laboratory, Hefei, Anhui 230088, China}
\affiliation{School of Emerging Technology, University of Science and Technology of China, Hefei, Anhui 230026, China}

\author{Hai-Zhou Lu}
\email{Corresponding author: luhz@sustech.edu.cn}
\affiliation{Department of Physics, State Key Laboratory of Quantum Functional Materials, and Guangdong Basic Research Center of Excellence for Quantum Science, Southern University of Science and Technology (SUSTech), Shenzhen 518055, China}
\affiliation{Quantum Science Center of Guangdong-Hong Kong-Macao Greater Bay Area (Guangdong), Shenzhen 518045, China}

\author{X. C. Xie}
\affiliation{Interdisciplinary Center for Theoretical Physics and Information Sciences (ICTPIS), Fudan University, Shanghai 200433, China}
\affiliation{International Center for Quantum Materials, School of Physics, Peking University, Beijing 100871, China}
\affiliation{Hefei National Laboratory, Hefei, Anhui 230088, China}

\date{\today }

\begin{abstract}
An altermagnet is a newly discovered antiferromagnet, characterized by unique anisotropic spin-split energy bands. It has attracted tremendous interest because of its promising potential in information storage and processing.   
However, measuring the distinctive spin-split energy bands arising from altermagnetism remains a challenge. Here, we propose to employ the Coulomb drag to probe altermagnetism. 
In the Coulomb drag, an electric current in an active layer of electron gases can induce currents in a close but well-isolated passive layer, due to interlayer Coulomb interactions. We find that the Coulomb drag effects in altermagnets are highly sensitive to the orientation of the spin-split Fermi surfaces. As a result, transverse currents can be dragged in the passive layer, leading to Hall drag effects even in the absence of spin-orbit coupling, a feature quite different from all previous systems. 
More importantly, all the drag effects of altermagnets have unique angle dependence, which can be measured in a multiterminal setup to serve as signatures for altermagnetism. This proposal will inspire increasing explorations on emergent magnetism.

\end{abstract}

\maketitle
\emph{{\color{blue}{Introduction.--}}} Altermagnetism is a newly recognized magnetic phase categorized by spin-group symmetries and features anisotropic spin-split bands that arise without spin-orbit coupling~\cite{hayami19JPSJ,Hayami20PRB,Libor22PRXLandscape,Libor22PRX2,Naka19NC,Ahn19PRB,yuanLD20PRB,Libor20SciAdv,ShaoDF21NC,2024arXiv240602123B,LiuJW21NC}. Recent studies have identified a growing number of candidate materials for altermagnets~\cite{YaQG23MTP,Mazin21PNAS,YaoYG24PRL,Osumi24PRB,Krempask24Nat,Lee24PRL,Moreno16PCCP,Marko242DMat,Reimers24NC,JianYD24arxiv,Naka20PRB,Das24PRL,JiangB24arxiv,FaYZ24arxiv,JiangB24arxiv,FaYZ24arxiv,Jiang2024arXiv}, including $\text{RuO}_2$~\cite{Ahn19PRB,ShaoDF21NC,Libor22PRX2,ZiHL24arxiv}, 
$\text{MnTe}$~\cite{Osumi24PRB,Krempask24Nat,Lee24PRL}, CrSb~\cite{Reimers24NC,JianYD24arxiv}, Rb$_{1-\delta}$V$_2$Te$_2$O~\cite{FaYZ24arxiv}, KV$_2$Se$_2$O~\cite{Jiang2024arXiv} and $\text{RuF}_4$~\cite{Marko242DMat}. One of the key challenges and focuses in this field is experimentally probing the distinctive Fermi surfaces and exploring their properties beyond conventional antiferromagnetism.~The anisotropic spin-split Fermi surfaces have been predicted to give rise to fascinating phenomena~\cite{DiZhu23PRB,YYG24PRL,WeiMM24PRB,FangY24PRL,Ouassou23PRL,Amundsen24PRB,SunC23PRB,PengY24PRL,HaiP24arxiv,ChenW24arxiv,ChenR24arxiv,Debashish24arxiv}, such as giant tunneling magnetoresistance~\cite{ShaoDF21NC,Libor22PRX1}, electrical spin splitter~\cite{LiuJW21NC,Rafael21PRL,Bose22NE}, and finite-momentum Cooper pairing~\cite{ZhangSB24NC,Sumita23PRB,2023arXiv230914427C,Sim2024arXiv,2024arXiv240702059H}, however, experimental verification of these Fermi surfaces remains limited. Recent efforts using transport, muon spin rotation, relaxation, and angle-resolved photoemission spectroscopy experiments have yielded controversial conclusions regarding altermagnetism in $\text{RuO}_2$~\cite{Hiraishi24PRL,Kessler24npj,Liu24PRL,ZHLin24arxiv}. Thus, innovative measurement techniques are crucial to unveil the true nature of altermagnets.

 \begin{figure}[t]
\includegraphics[width=1.0\linewidth]{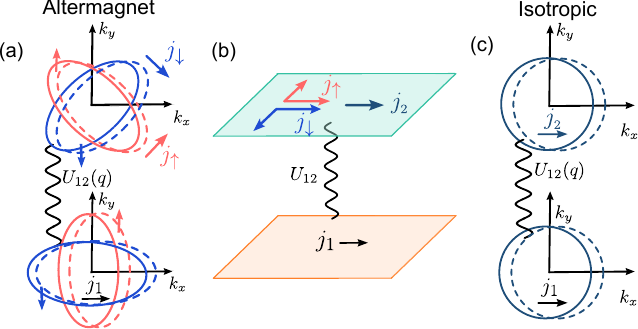}
\caption{Schematic of 
Coulomb drag setup. (a) Anisotropic spin-up (red) and spin-down (blue) Fermi surfaces in two isolated layers of altermagnet solely coupled by the interlayer Coulomb interaction $U_{12}(q)$. A driving current density $j_1$ shifts the Fermi surfaces, as indicated by the dashed rings. (b) In real space, $j_1$ applied in the active (bottom) layer drags currents in the passive (top) layer through $U_{12}$.  
The altermagnet uniquely supports both longitudinal and transverse currents ($j_{\uparrow}$ and $j_{\downarrow}$) relative to the transferred momentum $\bm q$ along the $j_1$ direction, leading to transverse Hall drag effects (Fig.~\ref{Fig:distinct_drags}) and angle dependences (Fig.~\ref{fig:angle_dependence}) even without spin-orbit coupling, thus giving signatures for altermagnetism. (c) In comparison, isotropic Fermi surfaces in conventional systems result in only a longitudinal dragged current ($j_2$) aligned with the transferred momentum. 
}\label{Fig:setup}
\end{figure}
\begin{figure*}[t]
\includegraphics[width=1.0\textwidth]{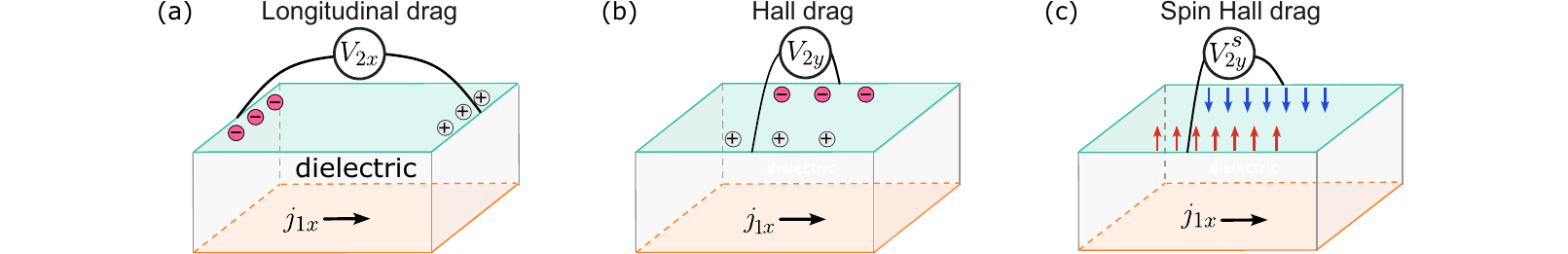}
\caption{Schematics of (a) the longitudinal Coulomb drag, (b) Hall drag, and (c) spin Hall drag effects
in altermagnet bilayers isolated by a dielectric layer. $j_{1x}$ is the injected electric current density in the active layer, while $V_{2x(y)}$ and $V_{2y}^{s}$ are the measured charge and spin voltages in the passive layer, respectively. Pink and gray circles represent negative and positive charges, respectively, while red and blue arrows in (c) indicate spins.
}
\label{Fig:distinct_drags}
\end{figure*}

\begin{figure*}[t] 
    \centering
    \includegraphics[width=1.0 \textwidth]{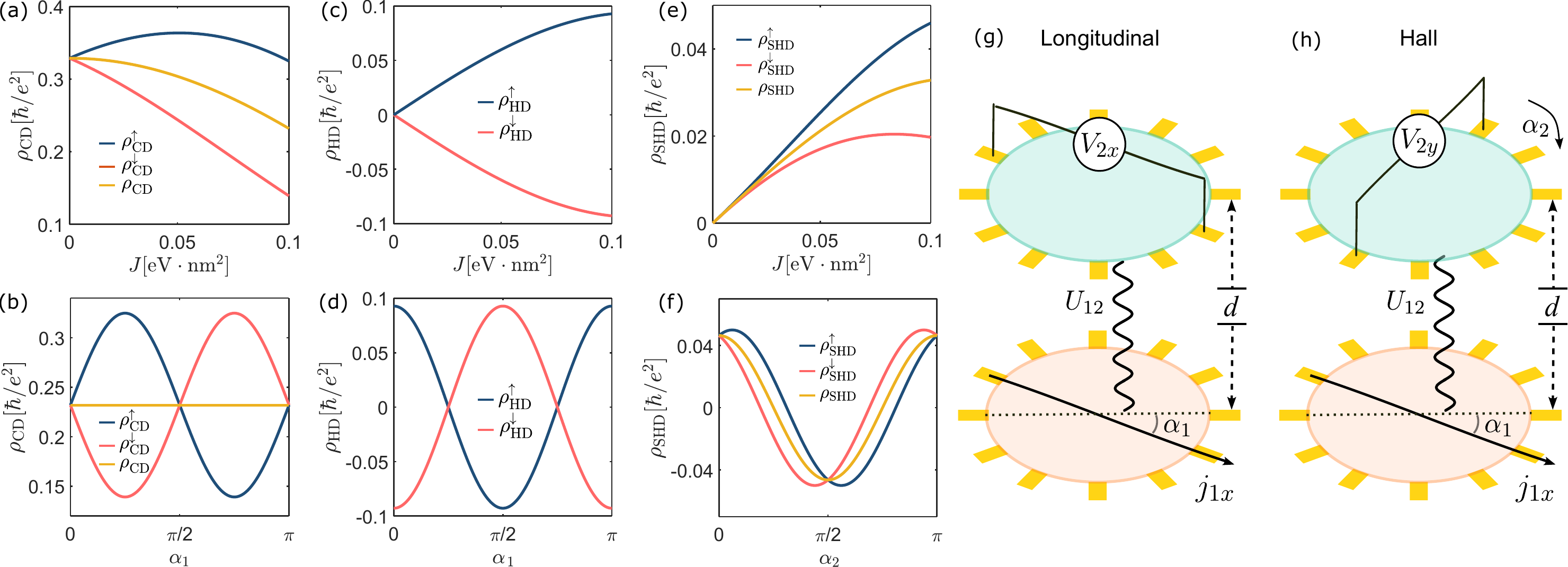} 
    \caption{(a),(b) Longitudinal drag resistivities $\rho_{\text{CD}}$, $\rho_{\text{CD}}^\uparrow$, and $\rho_{\text{CD}}^\downarrow$ for spin-unpolarized, spin-up and spin-down driving currents as functions of $J$ for $\alpha_1=\pi/4$ (a)
    and as functions of $\alpha_1$ for $J=0.8t$ (b). 
    (c),(d) Hall drag resistivities $\rho_{\text{HD}}^\uparrow$ and $\rho_{\text{HD}}^\downarrow$ for spin-up and spin-down polarized driving currents as functions of $J$ for $\alpha_1=\pi/2$ (c) and as functions of $\alpha_1$ for $J=0.8t$ (d). 
    (e),(f) Spin Hall resistivities $\rho_{\text{CD}}$, $\rho_{\text{CD}}^\uparrow$, and $\rho_{\text{CD}}^\downarrow$, for unpolarized, spin-up, and spin-down polarized driving currents as functions of $J$ for $\alpha_1=0$ and $\alpha_2=\pi/8$ (e) and as functions of $\alpha_2=\pi/8$ for $J=0.8t$ (f). 
(g),(h) Schematics of the longitudinal and Hall drag measurement setups of the altermagnetic bilayer with a dielectric of thickness $d$ in between and coupled by Coulomb interaction $U_{12}$. Equivalently, the angles $\alpha_1$ and $\alpha_2$ correspond to the orientations of the electrode pairs relative to the direction of the driving current $j_{1x}$. Other parameters are $d=12~ \text{nm}$, $T=2$ K, $\varepsilon_r=5$, $\tau=10^{-10} \text{s}$, $\mu_1=\mu_2= 0.3~\text{eV}$, and $t=0.127 ~ \text{eV}~ \text{nm}^2$ (corresponding to $m=0.3m_e$).
    }
\label{fig:angle_dependence}
\end{figure*}

In this Letter, we propose to use the Coulomb drag to explore altermagnetism. Coulomb drag involves two closely spaced, electrically isolated layers-an active layer, where a driving current is applied, and a passive layer, where currents induced via interlayer Coulomb interactions can be detected (Fig.~\ref{Fig:setup})~\cite{,Narozhny16RMP,Amico00PRB,Amico03PRB,Badalyan08PRL,Flensberg01PRB,Weber05Nat,Vignale05PRB,Badalyan09PRL,Driel10PRL}. Compared to the ordinary transport measurements, Coulomb drag allows noncontact detection of fundamental properties, such as electron-electron interaction~\cite{Laroche14Sci,Price07Sci,ZouY09PRB,Shimada00PRL}, exciton condensate~\cite{Li17NP,Nandi12Nat,Seamons09PRL,Mink12PRL,Eisenstein04Nat}, and 
quantum coherence~\cite{Zcg23NC,FengJ24Arxiv,Zcg23NP}. We study Coulomb drag in a $d$-wave altermagnetic bilayer. We find that the Coulomb drag resistivity highly depends on the orientation of the Fermi-surface splitting and spin polarization of the driving current. As a result, altermagnetism can give rise to Hall drag and spin Hall drag (Fig.~\ref{Fig:distinct_drags}), even in the absence of spin-orbit coupling, challenging the conventional knowledge that spin-orbit coupling is required for these transverse drag effects.
More importantly, all the Coulomb drag effects demonstrate angle dependence unique to altermagnets, which we propose to be measured in a multiterminal device (Fig.~\ref{fig:angle_dependence}). This Letter reveals unique signatures to identify altermagnetism and will inspire further explorations on novel magnetism.

 \textit{\color{blue}{Distinct Coulomb drag effects in altermagnets.-- }} In the Coulomb drag effect, a momentum of $\bm q$ is transferred from the active to the passive layer through interlayer Coulomb interactions. For conventional electron gases with isotropic Fermi surfaces [Fig.~\ref{Fig:setup}(c)], the momentum transferred from a driving current in the active layer only induces a drag current along the same direction in the passive layer. As a result, only the longitudinal Coulomb drag is allowed [Fig.~\ref{Fig:distinct_drags}(a)]~\cite{Onsager96PRL,Kam95PRB,JusticSong13PRL}.

The situation is fundamentally different in altermagnets with anisotropic spin-split energy bands [Fig.~\ref{Fig:setup}(a)], in particular when the Fermi-surface splittings of the active and passive layers have different orientations. The momentum transfer can result in drag currents not only in the parallel but also in transverse directions [Fig.~\ref{Fig:setup}(b)]. This behavior depends strongly on the relative direction between the driving current and the orientation of the Fermi-surface splitting. It enables not only an unusual longitudinal but also a Hall drag effect [Fig.~\ref{Fig:distinct_drags}(b)], even in the absence of spin-orbit coupling. 

Furthermore, in altermagnets, the two spin species exhibit distinct anisotropic band structures, which leads to independent responses to a driving current. This behavior is strongly dependent of the orientation of Fermi-surface spin splitting and the spin polarization of the driving current. Consequently, a spin Hall drag also emerges [Fig.~\ref{Fig:distinct_drags}(c)]. Better than the previous scenarios of spin-dependent Coulomb drag, in altermagnets, spin-polarized currents can be naturally generated in the active layer~\cite{Rafael21PRL,LiuJW21NC,Bose22NE}, without requiring sophisticated ferromagnetic electrodes.

All these unique Coulomb drag effects are intrinsically tied to the anisotropic spin-split Fermi surface, thus offering direct and distinctive signatures for detecting altermagnetism.

\begin{figure}[t]
\includegraphics[width=1.0\linewidth]{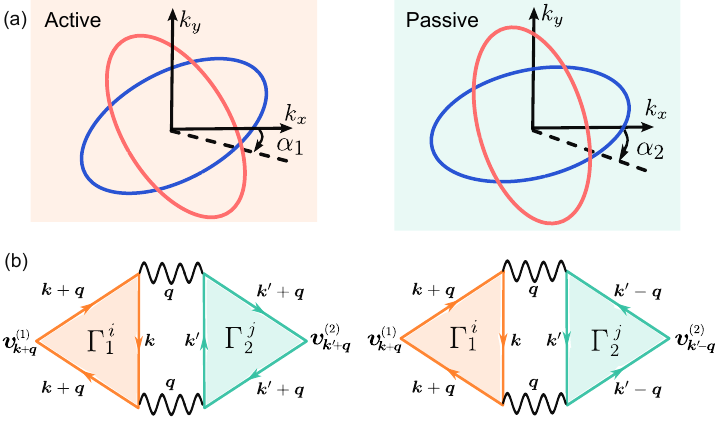}
\caption{(a) The angles $\alpha_1$ and $\alpha_2$ characterize the orientations of the altermagnetic spin splitting in the active and passive layers, respectively. (b) The Aslamazov–Larkin diagrams that give the Coulomb drag resistivity. The solid lines denote the single-particle Green functions and the wavy lines the Coulomb interaction. $\Gamma_{1(2)}^{i}$ is the nonlinear susceptibility of the active (passive) layer along the $i,j\in \{x,y\}$ direction.}
\label{Fig:dragram}
\end{figure}

\textit{\color{blue}{Model and formalism.-- }}To validate the above analysis, we consider two layers of $d$-wave altermagnet, separated by a dielectric layer of thickness $d$. In each layer, the two-dimensional (2D) $d$-wave altermagnet can be described by
the minimal model~\cite{ZhangSB24NC} 
\begin{equation}
\mathcal{H} (\bm{k})= tk^2 + J[\cos(2\alpha) k_xk_y + \sin(2\alpha)(k_x^2 -k_y^2)/2 ]\sigma_z, \label{Eq:model}
\end{equation}
where $k=|\bm{k}|$ and $\bm{k}=(k_x,k_y)$ is the wave vector. The $t$ term represents the kinetic energy, while the $J$ term captures the altermagnetic exchange interaction arising from the anisotropic crystal fields~\cite{Libor22PRXLandscape,Libor22PRX2}. $t=\hbar^2/(2m^*)$ is determined by the effective electron mass $m^*$. $\sigma_z$ is the $z$-direction Pauli matrix for spin. The angle $\alpha$ characterizes the orientation of the altermagnetic spin splitting, as illustrated in Fig.~\ref{Fig:dragram}(a) for the active ($\alpha_1$) and passive ($\alpha_2$) layers, respectively. The spin-split energy bands are given by $ E_{\bm{k}}^s =tk^2 +s J [\cos(2\alpha) k_xk_y + \sin(2\alpha)(k_x^2 -k_y^2)/2 ]$, where $s=+(-)$ denotes spin-up (down) state.
The model respects $[C_2||C_{4z}]$ symmetry-a fourfold spatial rotation about the $z$ axis combined with a spin flipping, imposing the $d$-wave magnetism with zero net magnetization. We consider the realistic scenario where $|J| < t$ and $m^*=0.3m_e$~\cite{Butkute98,Shai13PRL}, where $m_e$ is the free electron mass.

In the linear-response and weak-coupling regime, the drag resistivity can be calculated from the diagrams in Fig.~\ref{Fig:dragram} and found as~\cite{Narozhny07PRB,TseWK07PRBSOC,LiuH17PRB} 
\begin{align}
    \rho_D^{ij} &= \frac{e^2\hbar\beta}{16\pi \sigma_1\sigma_2 }\int \frac{d^2 q}{(2\pi)^2} \int_{-\infty}^\infty d\omega \frac{|U_{12}(q)|^2}{\sinh^2({\beta\omega}/{2})} \notag\\
    &\qquad \times\Gamma_{1}^{i}(\bm{q},\omega^+,\omega^-)\Gamma_{2}^{j}(\bm{q},\omega^-,\omega^+),
    \label{Eq:2}
\end{align}
where $i,j\in\{x,y\}$, $e$ is the elementary charge, $\hbar$ is the reduced Planck constant, and $\beta=1/(k_BT)$ with $T$ the temperature and $k_B$ the Boltzmann constant. The Drude conductivity of the active (passive) layer $\sigma_{1(2)} = 2te^2 \tau n_{1(2)}/\hbar^2$, with $n_{1(2)}$ the carrier density. At low temperatures ($k_BT$ $\ll$ $\mu_{1(2)}$), $n_{1(2)}$ can be approximated as $n_{1(2)}\approx \nu\mu_{1(2)}$, where $\mu_{1(2)}$ is the chemical potential and $\nu=(2\pi \sqrt{4t^2-J^2})^{-1}$ is the density of states (see \hyperlink{Appendix A}{Appendix A}). 
The Coulomb interaction 
$U_{12}({q}) = {e^2 q}/[{\varepsilon \kappa^2 \sinh (qd)}]$~\cite{Mahan,Narozhny16RMP} (see \hyperlink{Appendix B}{Appendix B}), under the random phase approximation and $q$ $\ll1/d$ in the Boltzmann regime, 
where $\varepsilon=\varepsilon_0\varepsilon_r$ with $\varepsilon_0$ the vacuum dielectric constant and $\varepsilon_r$ the relative dielectric constant of the material, and the inverse Thomas-Fermi screening length $\kappa=e^2\nu/\varepsilon$.

In Eq.~\eqref{Eq:2}, $\Gamma^i_{1(2)}$ is the $i$-direction component of the nonlinear susceptibility in the active (passive) layer, as shown by the triangle diagrams in Fig.~\ref{Fig:dragram}. $\omega$ is the frequency, and $\omega^{\pm}=\omega \pm i0^+$ with $0^+$ a positive infinitesimal. At leading order, the susceptibility can be found proportional to the imaginary part of the retarded polarization as 
\begin{align}
{\bm \Gamma}_{\ell} &(\bm{q},\omega^\pm,\omega^\mp) =-{2\tau} {\bm v}_{\bm{q}}^{(\ell)} \text{Im} [\Pi^R_{\ell}(\bm{q},\omega)]/\hbar,
\end{align}
where the active and passive layers are $\Gamma_1(\bm{q},\omega^+,\omega^-)$ and  $\Gamma_2(\bm{q},\omega^-,\omega^+)$, respectively. $\bm{\Gamma}_{\ell} \equiv (\Gamma_{\ell}^{x}, \Gamma_{\ell}^{y})$, $\ell\in\{1,2\}$ is the layer index, $\tau$ is the scattering time, and $\bm{v}_{\bm{q}}^{(\ell)} = \bm{v}_{\bm{k} +\bm{q}}^{(\ell)} - \bm{v}_{\bm{k}}^{(\ell)}$ represents the shift of electron velocity caused by $\bm q$. The velocity from spin-split energy bands distinguishes the spin species. 
The polarization reads
\begin{align}
\Pi^{R}_{\ell}({\bm q},\omega) = \frac{1+i\omega\tau /\hbar}{2\pi \sqrt{4t^2-J^2}}.
\label{eq:polarization}
\end{align}
When $J=0$, Eq.~\eqref{eq:polarization} reduces to the result of the 2D free electron gas~\cite{Kam95PRB,FlensbergK95PRB}. The presence of altermagnetism ($J\neq 0$) increases the polarization and enhances the nonlinear susceptibility. Because of the different anisotropy of the spin-up and spin-down electron pockets, the transferred momentum $\bm q$ generally induces $\bm{v}_{\bm{q}}^{(2)}$ with both longitudinal and transverse components relative to $\bm q$, giving rise to not only the longitudinal but also transverse Coulomb drag effects in Fig.~\ref{Fig:distinct_drags}.

Under the above considerations, the Coulomb drag resistivities can be derived analytically and summarized in Table~\ref{table:expressions} 
(see \hyperlink{Appendix C}{Appendix C}), where the common factor
\begin{align}
\mathcal{F}_T = \frac{5\zeta(5)\pi^4(k_BT)^2}{\hbar e^6d^6} \frac{\varepsilon^2\tau^2}{ \mu_1\mu_2} \frac{(4t^2-J^2)^2}{t^2}
\label{eq:factor}
\end{align}
depends quadratically on temperature and the scattering time. $\zeta(x)$ is the Riemann zeta function. As expected, all drag resistivities generally appear and depend significantly on the strength $J$ and both orientations $\alpha_{1(2)}$ of the altermagnetic spin splitting in the two layers. Note that the intertwined angle ($\alpha_{1(2)}$) dependences are absent in conventional electron gases even with spin-degenerate anisotropic Fermi surfaces. More importantly, the transverse components ($\rho_{\sigma\sigma'}^{xy}$, $\sigma, \sigma' \in \{\uparrow, \downarrow\}$) are directly linked to the altermagnetic strength $J$, becoming nonzero only in the presence of altermagnetism.

\begin{table}[t]
\centering
\caption{Drag resistivities $\rho^{ij}_{ss'}$ at low temperatures ($k_BT \ll \mu$), describing the drag voltage of spin $s^\prime$ along the $j$ direction in the passive layer driven by a 
current of spin $s$ along the $i$ direction in the active layer. 
Here, $\alpha_\pm =\alpha_1\pm \alpha_2$, and the angles $\alpha_1$ and $\alpha_2$ characterize the orientations of altermagnetic spin splitting in the active and passive layers, respectively, as illustrated in Figs. \ref{Fig:dragram}(a), \ref{fig:angle_dependence}g and \ref{fig:angle_dependence}(h).}
\setlength{\tabcolsep}{0.5cm}
\begin{tabular}{ccl}
    \hline
\hline
\text{Resistivities} & \text{Expressions} \\
\hline
    $\rho_{\uparrow\uparrow}^{xx}/\mathcal{F}_T$ \  &  $ 4t^2 +J^2 \cos 2\alpha_- + 4tJ \sin \alpha_+\cos \alpha_- $ \vspace{1mm} \\ 
   $\rho_{\downarrow\downarrow}^{xx}/\mathcal{F}_T$ \  & $ 4t^2 +J^2 \cos 2\alpha_- - 4tJ \sin \alpha_+ \cos \alpha_- $ \vspace{1mm} \\
 $\rho_{\uparrow\downarrow}^{xx}/\mathcal{F}_T$ \  & $ 4t^2  -J^2\cos 2\alpha_- +4tJ \cos\alpha_+ \sin \alpha_-$ \vspace{1mm} \\
    $\rho_{\downarrow\uparrow}^{xx}/\mathcal{F}_T$ \  & $ 4t^2  -J^2\cos2\alpha_- -4tJ \cos\alpha_+\sin \alpha_-$ \vspace{1mm} \\ 
    $\rho_{\uparrow\uparrow}^{xy}/\mathcal{F}_T$ \  & $2J \cos \alpha_+(2t\cos\alpha_- + J\sin \alpha_+)$ \vspace{1mm} \\
    $\rho_{\downarrow\downarrow}^{xy}/\mathcal{F}_T$ \  & $2J \cos \alpha_+
    (-2t\cos \alpha_- + J\sin \alpha_+)$ \vspace{1mm} \\
    $\rho_{\uparrow\downarrow}^{xy}/\mathcal{F}_T$ \  & $ - 2J \sin \alpha_+ ( 2t \sin \alpha_-+J\cos \alpha_+)$ \vspace{1mm} \\
    $\rho_{\downarrow\uparrow}^{xy}/\mathcal{F}_T$ \  & $ -2J \sin \alpha_+ (-2t \sin \alpha_- +J\cos \alpha_+)$ \vspace{0.5mm} \\
        \hline  \hline  \end{tabular}
\label{table:expressions}
\end{table}

\textit{\color{blue}{Orientation dependence in Coulomb drag.-- }}With the above drag resistivities in Table~\ref{table:expressions}, we now analyze three types of measurable drag effects (Fig.~\ref{Fig:distinct_drags}), driven by spin-unpolarized or spin-polarized driving currents in the active layer.
In altermagnets, an applied electric field can inject a current with spin polarization, depending on the relative angle of the electric field and the orientation of spin splitting~\cite{Rafael21PRL,LiuJW21NC}. 
As illustrated in Fig.~\ref{Fig:distinct_drags}(a), driven by a current along the $x$ direction in the active layer, the induced voltage in the passive layer along the same direction can be written as $V_{2x} = \rho_{\text{CD}}^{\uparrow} j_{1x,\uparrow} +  \rho_{\text{CD}}^{\downarrow} j_{1x,\downarrow}$. 
The drag resistivity for a fully polarized current ($j_{1x}=j_{1x,\uparrow}$ or $j_{1x,\downarrow}$) 
is given by $  
\rho_{\text{CD}}^{\uparrow(\downarrow)} = \rho_{\uparrow\uparrow(\downarrow\downarrow)}^{xx} + \rho_{\uparrow\downarrow(\downarrow\uparrow)}^{xx} =4\mathcal{F}_T [2t^2 \pm   tJ\sin(2\alpha_1)]$, which depends on $J$ and $\alpha_1$, as shown in Figs.~\ref{fig:angle_dependence}(a) and \ref{fig:angle_dependence}(b). Hence, $V_{2x}$ is found as
\begin{equation}
V_{2x} \propto  \mathcal{F}_{T}[2t^2 + \eta tJ\sin(2\alpha_1)],
\end{equation}
where $\eta = (j_{1x,\uparrow}-j_{1x,\downarrow})/j_{1x}$  defines the polarization of the driving current $j_{1x}=j_{1x,\uparrow}+j_{1x,\downarrow}$. An unpolarized current corresponds to $\eta=0$. In this situation, the only impact of nonzero $J$ is a reduction in the signal magnitude. Importantly, 
due to the presence of $J$ that measures altermagnetism, the voltage in the passive layer $V_{2x}$ shows a $\pi$ periodicity in $\alpha_1$,  giving a distinctive signature for altermagnetism.

\textit{\color{blue}{Hall drag and spin Hall drag.-- }}As illustrated in Figs.~\ref{Fig:distinct_drags}(b) and \ref{Fig:distinct_drags}(c), altermagnetism can induce both Hall drag (HD) and spin Hall drag (SHD) effects, even without spin-orbit coupling.
Driven by a current along the $x$ direction
in the active layer, the transverse voltage $V_{2y}$ measured along the $y$ direction in the passive layer 
 can be written as $V_{2y} = \rho_{\text{HD}}^{\uparrow} j_{1x,\uparrow}+ \rho_{\text{HD}}^{\downarrow}j_{1x,\downarrow}$. The fully spin-polarized resistivities are $
  \rho_{\text{HD}}^{\uparrow (\downarrow)} = \rho_{\uparrow\uparrow(\downarrow\downarrow)}^{xy} + \rho_{\uparrow\downarrow(\downarrow\uparrow)}^{xy} =\pm 4 t J \mathcal{F}_T \cos (2\alpha_1)$. These Hall resistivities increase monotonically as the altermagnetic strength $|J|$ ($<t$) increases [Fig.~\ref{fig:angle_dependence}(c)] and are opposite for opposite spins. Moreover, they are periodic in $\alpha_1$ with a period $\pi$ [Fig.~\ref{fig:angle_dependence}(d)]. 
Therefore, $V_{2y}$ is found as
\begin{eqnarray}
V_{2y}  \propto  \eta J \mathcal{F}_T \cos(2\alpha_1).
\end{eqnarray}
Notably, $V_{2y}$ vanishes in the absence of altermagnetism ($J=0$) or when the driving current is unpolarized ($\eta=0$).

The spin Hall drag, where a spin transverse voltage $V_{2y}^{s}$ is induced in the passive layer, can be driven by either unpolarized or polarized driving currents. Explicitly, the spin transverse voltage is given by $V_{2y}^{s}=\rho_{\text{SHD}}^{\uparrow} j_{1x,\uparrow}- \rho_{\text{SHD}}^{\downarrow} j_{1x,\downarrow}
$, where the Hall drag resistivities for fully spin-polarized currents are $\rho_{\text{{SHD}}}^{\uparrow(\downarrow)} = \rho_{\uparrow\uparrow(\downarrow\uparrow)}^{xy} - \rho_{\uparrow\downarrow(\downarrow\downarrow)}^{xy} =  J\mathcal{F}_T [2t \cos(2\alpha_2) \pm J \sin(2\alpha_1 + 2 \alpha_2)]$. The resistivities $\rho_{\text{{SHD}}}^{\uparrow(\downarrow)}$ depend strongly on the polarization of the driving current and the orientations ($\alpha_1$ and $\alpha_2$) of the Fermi-surface spin splitting, as illustrated in Figs.~\ref{fig:angle_dependence}(e) and \ref{fig:angle_dependence}(f).
Consequently, $V_{2y}^{s}$ can be expressed as
\begin{eqnarray}
  V_{2y}^{s} \propto J \mathcal{F}_T [\cos(2\alpha_1)+ \eta J \sin(2 \alpha_1+2\alpha_2)].
\end{eqnarray} 
 This shows that $V_{2y}^s$ displays a $\pi$ periodicity in $\alpha_{1}$ or $\alpha_{2}$, similar to the longitudinal drag effect.

\textit{\color{blue}{Experimental implementation.--}} 
$d$-wave planar altermagnets have been predicted in various compounds, including $\text{RuO}_2$~\cite{Ahn19PRB,ShaoDF21NC,Libor22PRX2,ZiHL24arxiv}, $\text{MnF}_2$~\cite{yuanLD20PRB} and $\text{RuF}_4$~\cite{Marko242DMat}. Thin films of  $\text{RuO}_2$ have also been realized experimentally~\cite{Uchida20PRL,Occhialini22PRM}. These materials provide a promising platform to test our predictions. While small but finite spin-orbit coupling may coexist in these systems, leading to spin mixing, it does not qualitatively alter our main results as long as its magnitude remains much smaller than the altermagnetic splitting (detailed in the Supplemental Material Sec.~SVI~\cite{supp}). The spin-dependent 
anisotropic nature in altermagnets gives rise to the unusual angle-dependent Coulomb drag effects, which are fundamentally different from conventional systems with spin-orbit coupling~\cite{Narozhny12PRB,TseWK07PRBSOC,Badalyan09PRL,Vignale09JPCM,LiuH17PRB}, where there is no such angle dependence. In experiments, the anisotropic nature of  altermagnets ensures that transferred momentum induces both longitudinal and transverse drags. Thus, the angle dependence from the passive layer persists, even when the active layer is replaced by a normal metal with isotropic Fermi surfaces. If a normal metal serves as the active layer, only the spin-polarized current in it can generate a transverse drag signal in the passive layer, while the unpolarized current can generate a transverse spin-drag signal. Although our work focuses on $d$-wave altermagnetism, our main results can extend to other forms of planar altermagnetism such as $g$-wave and $i$-wave, with corresponding periods of angle dependence of $\pi/2$ and $\pi/3$, respectively.

\begin{acknowledgments}
We thank Jinsong Xu, Xiangang Wan and Xiaoqun Wang for valuable discussions. This work was supported by the National Key R\&D Program of China (Grant No.~2022YFA1403700), Innovation Program for Quantum Science and Technology (2021ZD0302400), the National Natural Science Foundation of China (Grants No.~11925402 and 12350402), Guangdong province (Grant No.~2020KCXTD001), Guangdong Basic and Applied Basic Research Foundation (2023B0303000011), Guangdong Provincial Quantum Science Strategic Initiative (GDZX2201001 and GDZX2401001), the Science, Technology and Innovation Commission of Shenzhen Municipality (Grant No.~ZDSYS20190902092905285), and the New Cornerstone Science Foundation through the XPLORER PRIZE. S.B.Z. was supported by the start-up fund at HFNL, the Innovation Program for Quantum Science and Technology (Grant No. 2021ZD0302800). The numerical calculations were supported by Center for Computational Science and Engineering of SUSTech.

$^\dag$H.-J.L. and S.-B.Z. contributed equally.
\end{acknowledgments}

\bibliographystyle{apsrev4-1-etal-title_10authors}
\bibliography{ref}



\appendix 

\begin{figure*}[t]\label{Sec:End_Matter}
\centering
 \large \textbf{End Matter} 
\end{figure*}

\newpage

\renewcommand{\theequation}{A\arabic{equation}}
\setcounter{equation}{0}

\hypertarget{Appendix A}{{\color{blue}\emph{Appendix A: Derivation of density of states.}}}~At low temperatures, the density of states $\nu_s$ is derived from the eigenvalues $E_{\bm{k}}^s$ of the model in Eq.~\eqref{Eq:model}, using the formula 
\begin{equation}
    \nu_s = \int d^2 k/(2\pi)^2\delta(\epsilon - E_{\bm{k}}^s).
\end{equation}  
We first convert the integral to polar coordinates, defining $\phi =\arctan (k_y/k_x)$ and $k=|\bm{k}|$. After applying the delta function to perform the integration over $k$, $\nu_s$ becomes 
\begin{align}
\nu_s = \frac{1}{(2\pi)^2} \int_{0}^{2\pi} \frac{d\phi}{2t + sJ \sin 2\phi}.
\end{align}
Next, we use the substitution $z=e^{i\phi}$, which maps the angular integral onto a contour integral over the unit circle $|z|=1$:
\begin{align}
\nu_s = \frac{1}{(2\pi)^2} \oint_{|z|=1} \frac{2z dz}{4t i z^2 + sJ(z^4 - 1)}.
\end{align}
Finally, we apply the residue theorem to the poles inside the contour and find the density of states as
\begin{equation}
\nu_s = \frac{1}{2\pi\sqrt{4t^2 - J^2}},
\end{equation}
which is valid in the regime where $|J|<2t$.

\renewcommand{\theequation}{B\arabic{equation}}
\setcounter{equation}{0}

\hypertarget{Appendix B}{{\color{blue}\emph{Appendix B: Derivation of screened Coulomb interaction.}}}~To derive the screened Coulomb interaction, we first calculate the polarization operator, which is given by
\begin{align}
\Pi(\bm{q},i\omega_m) = \frac{1}{\beta S} \sum_{\bm{k}} \sum_{i\varepsilon_n}   G_{\bm{k+q}}^{i\epsilon_n+i\omega_m} G_{\bm{k}}^{i\varepsilon_n},
\end{align} 
where $\bm{q}$ is the transferred momentum, $i\omega_m$ and $i\epsilon_n$ are Matsubara frequencies for bosons and fermions, respectively. The Green functions are $G_{\bm{k}}^{i\epsilon_n} = (i\epsilon_n - E_{\bm{k}}^s)^{-1}$ and $G_{\bm{k}+\bm{q}}^{i\epsilon_n+i\omega_m} = (i\epsilon_n+i\omega_m - E_{\bm{k}+\bm{q}}^s)^{-1}$. To perform the Matsubara frequency summation and for convenience, we define the function 
$f(i\epsilon_n+i\omega_m,i\epsilon_n) \equiv G_{\bm{k+q}}^{i\epsilon_n+i\omega_m} G_{\bm{k}}^{i\varepsilon_n}$ and convert the summation into an integral, yielding
\begin{align}
\Pi(\bm{q},i\omega_m) = \frac{1}{ S } \sum_{\bm{k}} \oint_c \frac{dz}{2\pi i}n_F(z)  f(z+i\omega_m,z),
\end{align}
where $z=i\epsilon_n$, $n_F(z)=1/((1+\exp[(z-\mu_\ell)/k_BT])$ is the Fermi distribution function, $T$ and $k_B$ are the temperature and Boltzmann constant. The contour of integration encloses only the poles of $n_F(z)$~\cite{Kam95PRB}.  Applying the branch cuts method at poles $z_1=\epsilon$ and $z_2=\epsilon-i\omega_m$ and performing the analytic continuation $i\omega_{m}\rightarrow \omega+i0^+$, we decouple the polarization into two parts: $   \Pi(\bm{q},i\omega_m)=\Pi_a^R(\bm{q},\omega) +\Pi_b^R(\bm{q},\omega)$, 
where 
\begin{subequations}
 \begin{align}
       \Pi_a^R &=\frac{1}{ S } \sum_{\bm{k}} \int_{-\infty}^{\infty} \frac{d\epsilon}{2\pi i}  n_F(\epsilon) f_1, \label{Eq:pi_a} \\
        \Pi_b^R &=\frac{1}{ S } \sum_{\bm{k}} \int_{-\infty}^{\infty} \frac{d\epsilon}{2\pi i}  [n_F(\epsilon+\omega)-n_F(\epsilon)]f_2,\label{Eq:pi_b}
\end{align}
\end{subequations}
Here, $f_1=f(\epsilon^++\omega,\epsilon^+)-f(\epsilon^-,\epsilon^- - \omega) $, $f_2 =f(\epsilon^++\omega, \epsilon^-)$, $\epsilon^\pm= \epsilon \pm i0^+$, and $0^+$ is a positive infinitesimal. The first part of the retarded polarization operator, $\Pi_a^R$, is found as 
\begin{align}
    \Pi_a^R(\bm{q},\omega) &= \int_{-\infty}^{\infty} \frac{d\epsilon}{2\pi i} \int \frac{d^2k}{(2\pi)^2} n_F(\epsilon)  \notag\\
    &\quad \times ( G_{\bm{k}+\bm{q},\epsilon+\omega}^R G_{\bm{k},\epsilon}^R - G_{\bm{k}+\bm{q},\epsilon}^AG_{\bm{k},\epsilon-\omega}^A ).  
\end{align} 
In the long-wavelength limit ($\bm{q}\rightarrow 0$), the leading order term can be related to the density of states~\cite{FlensbergK95PRB} by
\begin{align}
    \Pi_a^R(\bm{q},\omega) \approx \int_{-\infty}^\infty \frac{d\epsilon}{2\pi i} \frac{\partial n_F(\epsilon)}{\partial \epsilon} \int \frac{d^2k}{(2\pi)^2} (G^R_{\bm{k},\epsilon}-G^A_{\bm{k},\epsilon}), 
\end{align}
Using the relation between the Green functions and spectral function $A({\bf k},\epsilon)$, i.e., $i(G^R_{\bm{k},\epsilon}-G^A_{\bm{k},\epsilon})=A(\bm{k},\epsilon)$, considering the clean limit and substituting the density of states~\cite{TseWK07PRBSOC}, we obtain $\Pi_a^R=(2\pi\sqrt{4t^2-J^2})^{-1}$. 

Similar, for the second part of the polarization operator, Eq.~\eqref{Eq:pi_b}, we have
\begin{align}
\Pi_b^R({\bm q},\omega) 
&=\int \frac{d^2k}{(2\pi)^2} \int_{-\infty}^{\infty} \frac{d\epsilon}{2\pi i} n_F(\epsilon) \notag\\
&\quad \times \left( G_{\bm{k}+\bm{q},\epsilon}^R G_{\bm{k},\epsilon-\omega}^A - G_{\bm{k}+\bm{q},\epsilon+\omega}^R G_{\bm{k},\epsilon}^A  \right),
\end{align}
Expanding to the lowest non-vanishing order, we find 
\begin{align}
     G_{\bm{k}+\bm{q},\epsilon}^R G_{\bm{k},\epsilon-\omega}^A - G_{\bm{k}+\bm{q},\epsilon+\omega}^R G_{\bm{k},\epsilon}^A \approx  -\frac{\omega\tau}{\hbar}  \frac{\partial A(\bm{k},\epsilon)}{\partial \epsilon},
\end{align}
where $\tau$ is the scattering time.
Thus, the second part is $\Pi_b^R({\bm q},\omega)  
\approx i\omega\tau(2\pi\hbar\sqrt{4t^2-J^2})^{-1}$. Considering $\Pi^R_a$ and $\Pi_b^R$ together, the total polarization operator is 
\begin{align} \label{Eq:B2}
\Pi^R({\bm q},\omega) = \frac{1+i\omega\tau/\hbar}{2\pi\sqrt{4t^2-J^2}},
\end{align}

Under the random phase approximation, the screened interlayer potential can be written in terms of the polarization functions as
\begin{equation}
\label{Eq:B3}
U_{12} = \frac{u_{12}}{(1+u_1\Pi_1)(1+u_2\Pi_2)-u_{12}u_{21}\Pi_1\Pi_2},
\end{equation}
where $u_{12}=u_{21}=4\pi u_{1(2)} e^{-qd}$ is the bare interlayer Coulomb potential, with $d$ the interlayer dielectric thickness and the identical $u_{1}=u_2=e^2/(2\varepsilon q)$ the bare intralayer Coulomb potential for the active and passive layers, respectively. Here, $\varepsilon$ is the dielectric constant of material, $\Pi_\ell$ is the retarded polarization operator of the $\ell$-th layer in Eq.~\eqref{Eq:B2}.  In the altermagnetic bilayer, $\Pi_1=\Pi_2=\Pi$ are identical. Plugging the retarded polarization operator and bare Coulomb interaction into Eq.~\eqref{Eq:B3} and considering $\kappa d \gg 1$, we derive the screened interlayer Coulomb potential as presented in the main text.

\renewcommand{\thetable}{C\arabic{table}}
\renewcommand{\theequation}{C\arabic{equation}}
\setcounter{equation}{0}
\setcounter{table}{0}

\hypertarget{Appendix C}{{\color{blue}\emph{Appendix C: Derivation of drag resistivities in Table~\ref{table:expressions}.}}}~
To derive the drag resistivity, we first deal with the nonlinear susceptibility, which can be expressed as~\cite{FlensbergK95PRB} 
\begin{align}\label{Eq:C1}
& \bm{\Gamma}_{\ell}(\bm{q},\omega^\pm,\omega^\mp) \notag \\
= &-\frac{2\tau}{\hbar}  \text{Im} \left\{  \int_{-\infty}^{\infty} \frac{d\epsilon}{2\pi i }  \left[n_F(\epsilon+\omega)-n_F(\epsilon)\right] \right. \notag\\
&\qquad  \left. \times   \int \frac{d^2 k}{(2\pi)^2} [v_{\bm{k}+\bm{q}}^{(\ell)}  -  v_{\bm{k}}^{(\ell)}] G^R_{\bm{k}+\bm{q},\epsilon+\omega}G^A_{\bm{k},\epsilon} \right\}.
\end{align}
Here, the nonlinear susceptibilities for the active layer and passive layer are $\bm{\Gamma}_{1}(\bm{q},\omega^+,\omega^-)$ and $\bm{\Gamma}_{2}(\bm{q},\omega^-,\omega^+)$, respectively. They can be adapted to each other by replacing the velocity in each individual layer. For the altermagnetic metal with quadratic dispersion, the velocity difference $ v_{\bm{q}}^{(\ell)}= v_{\bm{k}+\bm{q}}^{(\ell)}  -  v_{\bm{k}}^{(\ell)}$ depends only on the transferred momentum $\bm q$,
akin to that of free electron gases~\cite{FlensbergK95PRB,Kam95PRB}. Explicitly, we find  
\begin{subequations}
\label{eq:velocity}
\begin{align}
v_{\bm{q},x}^{(\ell)} &= \frac{1}{\hbar} [ 2t q_x + sJ \cos(2\alpha_\ell) q_y + sJ \sin(2\alpha_\ell) q_x ],\\
v_{\bm{q},y}^{(\ell)}  &= \frac{1}{\hbar} [ 2t q_y + sJ \cos(2\alpha_\ell) q_x + sJ \sin(2\alpha_\ell) q_y ],
\end{align}
\end{subequations} 
where $s=\pm$ indicates the electron spin. Consequently, the nonlinear susceptibility can be found proportional to the imaginary part of the polarization operator~\cite{supp}:
\begin{align}
{\bm \Gamma_{\ell}}(\bm{q},\omega^\pm,\omega^\mp)
&=-2 \frac{\tau}{\hbar} \bm{v}_{\bm q}^{(\ell)} \text{Im} [\Pi^R({\bm q},\omega)].
\label{Eq:C2}
\end{align}

In the linear-response and weak-coupling regime, the drag resistivity can be formulated as
\begin{align}\label{Eq:C1}
    \rho_D^{ij} &= \frac{e^2\hbar\beta}{16\pi \sigma_1\sigma_2 }\int \frac{d^2 q}{(2\pi)^2} \int_{-\infty}^\infty d\omega \frac{|U_{12}(q)|^2}{\sinh^2({\beta\omega}/{2})} \notag\\
    &\qquad \times\Gamma_{1}^{i}(\bm{q},\omega^+,\omega^-)\Gamma_{2}^{j}(\bm{q},\omega^-,\omega^+),
\end{align}
where the superscripts $i,j\in\{x,y\}$. At low temperatures, the Drude conductivity of each layer is given by $\sigma_{1(2)} = 2e^2 t \nu \tau \mu_{1(2)}/\hbar^2$. Using the nonlinear susceptibility in Eq.~\eqref{Eq:C2}, the screened Coulomb potential in Eq.~\eqref{Eq:B3}, and the following identity, 
\begin{align} 
    \int_{-\infty}^{\infty}dx \frac{x^2}{\sinh^2(yx/2)} = \frac{8\pi^2}{3y^3},
\end{align}
to handle the frequency $\omega$ integral~\cite{Carrega12NJP,Narozhny07PRB},
the drag resistivity becomes
\begin{align}
\rho^{ij}_D&=  (k_BT)^2 \frac{2\pi^3\varepsilon^2 \tau^2\hbar }{3e^6\mu_1\mu_2}\frac{(4t^2-J^2)^2}{t^2}  \notag\\
&\quad \times \int_{0}^{\infty} q dq  \int_{0}^{2\pi} d\varphi \frac{q^2}{\sinh^2 (qd)}  {v}_{\bm{q}}^{(1)i} {v}_{\bm{q}}^{(2)j}.
\end{align}
Finally, by substituting the results in Eq.~\eqref{eq:velocity} for the group velocities and performing the integrals, we obtain the final results as presented in Table~\ref{table:expressions}.

We also find a spin Coulomb drag effect in an altermagnet monolayer in Sec.~SIV~\cite{supp}.

\renewcommand{\theequation}{D\arabic{equation}}
\setcounter{equation}{0}

\end{document}